
\magnification=1200
\parindent 1.0truecm
\baselineskip=16pt
\hsize=6.0truein
\vsize=8.5truein
\rm
\null

\footline={\hfil}
\vglue 0.8truecm
\rightline{\bf DFUPG-54-92}
\rightline{\sl May 1992 }
\vglue 2.0truecm
\centerline{\bf Low--Energy $KN$ Interactions at a $\phi$--Factory }
\vglue 2.0truecm
\centerline{\sl Paolo M. Gensini }
\centerline{\sl Dip. di Fisica dell\'\ Universit\`a di Perugia, Perugia,
Italy, and }
\centerline{\sl I.N.F.N., Sezione di Perugia, Italy. }
\vglue 5.0truecm
\centerline{ Talk presented at the }
\centerline{\sl VII Scuola Invernale di Fisica Adronica, }
\centerline{\sl Folgaria (Trento), Italy, February 10--15, 1992 }
\centerline{ To be published in the {\sl Proceedings, }}
\centerline{ ed. by T. Bressani, {\sl et al.} (World Scientific, Singapore
1992). }
\pageno=0
\vfill
\eject

\footline={\hss\tenrm\folio\hss}
\vglue 1.0truecm
\centerline{\bf Low--Energy $KN$ Interactions at a $\phi$--Factory }
\vglue 0.5truecm
\centerline{\sl Paolo M. Gensini }
\centerline{\sl Dip. di Fisica dell\'\ Universit\`a di Perugia, Perugia,
Italy, and }
\centerline{\sl I.N.F.N., Sezione di Perugia, Italy. }

\vglue 1.0truecm
\noindent{\bf 1. Introduction.}

We intend to illustrate in this lecture the
possibilities opening up at machines planned for the nineties for
low--energy kaon--nucleon interactions,
keeping the focus on the theoretical problems they should help to
solve.

We shall deal here only with the $\phi$--factory DA$\Phi$NE$^1$ and
consider, for brevity,
interactions with light, gaseous targets, using gaseous $H_2$ as a
benchmark for which to estimate the rates to be expected in a typical
apparatus.

The interest in this field is of a systematic rather than exploratory
nature: information on low--energy kaon--nucleon interactions is
scarce and of a poor statistical quality, when compared to the corresponding
pion--nucleon one. As an example, just take a look at the two pages
dedicated by the PDG booklet$^2$ to $K^\pm p$ and $K^\pm d$ total and elastic
cross sections: other data do not present a rosier perspective$^3$.

The low quality of low--momentum elastic and inelastic scattering data
reflects in turn on our knowledge of the ``elementary'' parameters of the
$K N$ interaction, remarkably poorer than in the $SU(3)_f$--related $\pi N$
case$^3$. On top of this sorry situation, one must add the still unsolved
mystery of kaonic--hydrogen level--shifts and widths, whose experimental
determinations are in total disagreement with theoretical expectations$^4$.

Data at very low momenta and at rest are essential to clarify many of
the above--mentioned problems$^3$; however, experiments of this kind pose
formidable problems at conventional fixed--targed machines, some of which can
be circumvented at a $\phi$--factory. For
instance, at the KAON factory planned for TRIUMF$^5$, beams in the lowest
momentum range (from 400 to 800 $MeV/c$) have intensities of $10^6$ -- $10^8$
$K^- s^{-1}$, with $K^+$ beams about twice more
intense. Already the purity of these beams is limited by $K^\pm$ decays
in flight: to experiment at momenta below 400 $MeV/c$ one has to use
moderators, which at the same time decrease the kaon intensity, degrade the
beam resolution, and increase enormously the beam contamination at the final
target. All these effects make the experiments much more complex, overturning
all the advantages offered by the higher initial beam intensities.

\vglue 0.6truecm
\noindent{\bf 2. Capabilities of a $KN$--scattering experiment at DA$\Phi$NE. }

DA$\Phi$NE is the $\phi$--factory (the acronym stands for ``Double Annular
$\Phi$--factory for Nice Experiments''), due to replace the Adone
colliding--beam machine in the same experimental hall of the I.N.F.N. National
Laboratories in Frascati. From its expected commissioning luminosity$^6$ of
$5 \times 10^{32}\ cm^{-2} s^{-1}$, and an annihilation cross section of about
5 $\mu b$ at the $\phi$--resonance peak, one can see that its {\sl two}
interaction regions will be the sources of $\simeq 1.2 \times 10^3$ $K^\pm
s^{-1}$, at a central momentum of 126.9 $MeV/c$, with the momentum resolution
of $\simeq 1.1 \times 10^{-2}$ due to the very small energy spreads in the
beams, as well as of $\simeq$ 850 $K_L s^{-1}$, at a central momentum of 110.1
$MeV/c$, with the slightly poorer resolution of $\simeq 1.5 \times 10^{-2}$.

Both $\pi^\pm$'s and leptons coming out of the two sources are
easy--to--control backgrounds: the first because the $\pi^\pm$'s, though
produced at a rate of about 380 $\pi^\pm s^{-1}$ (not counting those from
$K_S$ decays), come {\sl almost all} from events with three or more final
particles and can thus be suppressed by momentum and collinearity cuts; the
second, as well as collinear pions from $e^+ e^- \to \pi^+ \pi^-$, produced at
much lower rates of order 0.75 $s^{-1}$ (the leptons) or 0.25 $s^{-1}$ (the
pions), are eliminated by a momentum cut, having momenta about four times
those of the $K^\pm$'s.

The {\sl two} interaction regions are therefore {\sl small--sized} sources of
{\sl low--momen-tum, tagged} $K^\pm$'s and $K_L$'s, with {\sl negligible
contaminations} (after suitable cuts on angles and momenta of the particles
are applied {\sl event by event}), in an environment of very low background
radioactivity: this situation is simply unattainable with {\sl conventional}
technologies at fixed--target machines$^7$, where the impossibility of placing
experiments too close to the production target limits from below the
charged--kaon momenta, and kaon decays in flight always contaminate the beams:
low--momentum experiments are thus possible only with the use of moderators,
with a huge beam contamination at the target, as well as a large
final--momentum spread due to straggling phenomena.

It is therefore of interest to consider the feasibility of low--energy,
$K^\pm N$ and $K_L N$ experiments at DA$\Phi$NE, with respect to equivalent
projects at machines such as, {\sl e.g.}, KAON at TRIUMF$^8$ (or to ideas
advanced for the sadly aborted EHF project$^9$).

We shall, in this part, try and give an evaluation of rates to be expected
in a simple apparatus at DA$\Phi$NE. We shall assume cylindrical symmetry,
with a toroidal target fiducial volume, limited by radii $a$ and $a + d$ and
of length $L$ (inside and outside of which one can imagine a tracking system,
surrounded on the outside by a photon detecting system -- {\sl e.g}
lead--Sci--Fi sandwiches -- and a superconducting, solenoidal coil to provide
the moderate magnetic field {\bf B} needed for momentum measurements), filled
with a gas at moderate pressure.

One must convert the usual, fixed--target expression for reaction rates to a
spherical geometry, and also include kaon decays in flight, getting (for
{\bf B} = 0 or $K^0_L$'s: the general case can be easily treated with slight
modifications)
$$
d N_r = [ {1 \over \rho^2} ({3 \over {8 \pi}})\ (L \sigma_\phi B_\phi)
\sin^2 \theta e^{- \rho / \lambda} ] \sigma_r \rho_t (\rho^2 d \rho \sin
\theta d \theta d \phi)\ , \eqno (1)
$$
with $\rho$, $\theta$ and $\phi$ spherical coordinates (with the $z$--axis
oriented along the beam direction), $L$ the machine luminosity, $\sigma_\phi$
the annihilation cross section at the $\phi$--resonance peak, $B_\phi$ the
$\phi$ branching ratio into the desired mode (either $K^+ K^-$ or $K_L K_S$),
$\sigma_r$ the reaction cross section for the process considered, $\rho_t$ the
target {\sl nuclear} density, and $\lambda = p_K \tau_K / m_K$ the decay
length (respectively of 0.954 $m$ for $K^\pm$'s and of 3.429 $m$ for $K_L$'s)
at the $\phi$--resonance momenta.

Integrating over the fiducial volume, the reaction rate can be cast into the
simple formula
$$
N_r = {3 \pi \over 4} r d (L \sigma_\phi B_\phi) \rho_t \sigma_r \ , \eqno (2)
$$
with both geometrical acceptance and kaon decay in flight thrown into the
reduction factor $r$, which we have estimated to take the values 0.50 for
$K^\pm$'s and 0.72 for $K_L$'s for a fiducial volume defined by $a =$ 10 $cm$,
$d =$ 50 $cm$ and $L =$ 1 $m$, to represent a ``person--sized'' detector,
fitting in DA$\Phi$NE's interaction region. A longer detector or a larger
outer radius would not give any substantial improvement in the rates, due to
the angular distribution of the produced kaons and to the value of the decay
length $\lambda$ for these low--momentum $K$'s; besides, for $K^{\pm}$ $r$
increases almost linearly but slowly with increasing field {\bf B}, due to the
interplay of the increased path length inside the fiducial volume on one side,
and of the particle decays on the other.

This gives, for a target volume filled by a diatomic, ideal gas at room
temperature, the rates for $K^\pm$--initiated processes
$$
N_r = p(\hbox{atm}) \times \sigma_r(\hbox{mb}) \times (4.0\times 10^4\
\hbox{events/y}) \ , \eqno (3)
$$
for a ``Snowmass year'' of $10^7$ s (for $K_L$'s the figure in eq. (3) is
about the same, because of an approximate compensation between the variations
in $r$ and $B_\phi$), or, with rough estimates of the partial $K^- p$ cross
sections at the $\phi$--decay momenta, to about $10^7$ two--body events per
year in $H_2$ gas at atmospheric pressure, of which about $3.6 \times 10^6$
elastic scattering events, $2.4 \times 10^6$ $\pi^+ \Sigma^-$ and about $10^6$
for each of the remaining four two--body channels $\pi^0 \Sigma^0$, $\pi^0
\Lambda$, $\bar K^0 n$, and $\pi^- \Sigma^+$. The above rates are enough to
measure angular distributions in all channels, and also the polarizations for
the self--analyzing final--hyperon states, particularly for the decays
$\Lambda \to \pi^- p$, $\pi^0 n$ (asymmetry $\alpha \simeq 0.64$) and
$\Sigma^+ \to \pi^0 p$ ($\alpha \simeq -0.98$). One could also expect a total
of about $10^4$ radiative--capture events, which should allow a good
measurement on the {\sl absolute} rates for these processes as well.

Such an apparatus will need: tracking for incoming and outgoing charged
particles, time--of--flight measurements (for charged--particle
identification), a moderate magnetic field (due to the low momenta involved)
for momentum measurements, and a system of converters plus scintillators for
photon detection and subsequent geometrical reconstruction of $\pi^0$ and
$\Sigma^0$ decays, amounting thus to a rather simple (on today's
particle--physics scale), not too costly apparatus. Mentioning costs, we wish
to point out that DA$\Phi$NE, though giving the experimenters a very small
momentum range, saves them the cost of the {\sl separate} tagging system
needed to reject contaminations in a {\sl conventional} low--energy,
fixed--target experiment$^7$.

The above formul\ae\ for $K^\pm$ rates do not include particle losses in the
beam--pipe wall and in the internal tracking system, which were assumed
sufficiently thin ({\sl e. g.} of a few hundred $\mu m$ of low--$Z$ material,
such as carbon fibers or Mylar). We have indeed checked that, due to the shape
of the angular distribution of the kaons, particle losses are contained
(mostly at small angles, where $K$--production is negligible, and events would
anyhow be hard to be fully reconstructed), and momentum losses flat around
$\theta = \pi / 2$ (where most of the $K^\pm$'s are produced): even for a
total thickness of the above--mentioned materials of 1 $mm$, kaon momenta do
not decrease below 100 $MeV/c$ and losses do not grow beyond a few percents.
Rather, one could exploit such a thickness as a low--momentum, thin moderator,
to span the interesting region just above the charge--exchange threshold at
$p_L(K^-) \simeq$ 90 $MeV/c$, measurements which would add precious,
additional constraints on low--energy amplitude analyses$^{10}$.

We have presented the above simplified estimates to show that acceptable rates
can be achieved, orders of magnitude above those of existing data at about the
same momentum, {\sl i.e.} to the lowest--energy points of the British--Polish
Track--Sensitive Target (TST) Collaboration, taken in the mid and late
seventies at the (R.I.P.) NIMROD accelerator$^{11}$.

Since losses do not affect $K_L$'s, a detector of the kind sketched above,
similar in geometry to the one proposed by T. Bressani$^7$ to do
$K^+$--nucleus scattering and hypernuclear experiments, could be used without
any problem to study low--energy $K_L \to K_S$ regeneration and
charge--exchange in gaseous targets, providing essential
information for this kind of phenomena.

We wish to add that a DA$\Phi$NE detector dedicated to kaon experiments on
gaseous $H_2$ and $D_2$ can continue its active life, without substantial
changes, to measure $K^+$--, $K^-$--, and $K^0_L$--interactions on heavier
gases as well ($He$, $N_2$, $O_2$, $Ne$, $Ar$, $Kr$, $Xe$), exploring not
only the properly nuclear aspects of these interactions, such as nucleon
swelling in nuclei$^{12}$, but also producing $\pi \Sigma$, $\pi \Lambda$ and
$\pi \pi \Lambda$ systems at invariant masses below the elastic $\bar K N$
threshold in the so--called unphysical region, with statistics substantially
higher than those now available$^{13}$, due to the $\simeq 4 \pi$ geometry
allowed by a colliding--beam--machine detector.

\vfill\eject
\vglue 0.6truecm
\noindent{\bf 3. $KN$ and $\bar KN$ amplitudes at low momenta: formalism and
some problems. }

Any $a_1(0^-,q) + B_1({1\over2}^+,p) \to a_2(0^-,q') + B_2({1\over2}^+,p')$
process is described in the c.m. frame by two amplitudes $G(w,\theta)$ and
$H(w,\theta)$, when the T--matrix element $T_{\alpha\beta}$ is expressed in
terms of two--component Pauli spinors $\chi_\alpha$ and $\chi_\beta$ for the
final and initial ${1\over2}^+$ baryon as $T_{\alpha\beta} =
\chi_\alpha^{\dag} \hbox{\bf T} \chi_\beta$, where
$$ \hbox{\bf T} = G(w,\theta) \times \hbox{\bf I} + i H(w,\theta) \times
(\vec \sigma \cdot \hat n) \eqno(4) $$
and $\hat n$ defines the normal to the scattering plane$^{14}$.

These amplitudes have a simple partial--wave expansion, given by
$$ G_N(w,\theta) = \sum_{\ell=0}^\infty [ (\ell + 1) T_{\ell+}(w) + \ell \
T_{\ell-}(w) ] P_\ell(\cos\theta) \eqno(5) $$
and
$$ H_N(w,\theta) = \sum_{\ell=1}^\infty [ T_{\ell+}(w) - T_{\ell-}(w) ]
P'_\ell(\cos\theta)\ , \eqno(6) $$
where the subscript $N$ indicates that only the nuclear interaction has been
considered. To describe the data, they must include electromagnetism and be
rewritten as
$$ G(w,\theta) = \tilde G_N(w,\theta) + G_C(w,\theta) \eqno(7)$$
and
$$ H(w,\theta) = \tilde H_N(w,\theta) + H_C(w,\theta) \ ,\eqno(8)$$
where the tilded amplitudes differ from the untilded ones in the Coulomb
shifts $\sigma_{\ell\pm}^{\rm in(fin)}$, non--zero only for {\sl both} initial
(final) charged particles, having been applied to each partial wave:
$$ T_{\ell\pm}\ \to \ \tilde T_{\ell\pm} = e^{i \sigma_{\ell\pm}^{\rm
in}} T_{\ell\pm}(w) e^{i \sigma_{\ell\pm}^{\rm fin}} \ . \eqno(9)$$

The one--photon--exchange amplitudes $G_C$ and $H_C$ (absent for charge--
or strangeness--exchange processes, but not for $K_S$ regeneration, which
at $t \neq 0$ goes also via one--photon exchange) are expressed in terms of
the nucleon Dirac form factors as$^{14,15}$ ($\mu$ and $m$ indicate
respectively the meson and baryon masses)
$$ G_C(w,\theta) = \pm e^{\pm i \phi_C} \cdot \{ ( {{2 q \gamma} \over t}
+ {\alpha \over {2 w}} {{w + m} \over {E + m}} ) \cdot F_1(t) + $$
$$ + [ w - m + {t
\over {4 (E + m)}} ] \cdot {{\alpha F_2(t)} \over {2 w m}} \} \cdot F_K(t)
\eqno(10) $$
and
$$ H_C(w,\theta) = \pm {{\alpha F_K(t)} \over {2 w \tan {1\over2} \theta}}
\cdot \{ {{w + m} \over {E + m}} \cdot F_1(t) + [w + {t \over {4 (E + m)}}]
\cdot {{F_2(t)} \over m} \} \eqno(11) $$
for $K^\pm$ interactions with nucleons, while for $K_S$ regeneration one has
to change the sign of the isovector part of the kaon form factor $F_K(t)$.

Here $\gamma = \alpha \cdot (w^2 - m^2 - \mu^2) / 2 q w$ and the Coulomb phase
$\phi_C$ is defined as
$$ \phi_C = - \gamma \log(\sin^2{1\over2}\theta) + \gamma \cdot \int_{-4q^2}^0
{{dt} \over t} \cdot [1 - F_K(t) F_1(t)] \eqno(12) $$
for charged kaons scattering on protons, while it reduces to
$$ \phi_C = -\gamma \int_{-4q^2}^0 {{dt}\over t} F_K(t) F_1(t) \eqno(12') $$
for processes involving $K^0$'s and/or neutrons.

$w$ and $\theta$ are the c.m. total energy and scattering angle, $q = [{1
\over 4} w^2 - {1 \over 2} (m^2 + \mu^2) + (m^2 - \mu^2)^2 / 4 w^2]^{1/2}$ the
c.m. momentum (in the initial state: for inelastic processes, including charge
exchange, we indicate final--state quantities with primes), $E$ the c.m.
(initial) baryon energy $E = (w^2 + m^2 - \mu^2)/2w$, and $t$ the squared
momentum--transfer, $t = m^2 + m'^2 - 2 E E' + 2 q q' \cos\theta$. We shall
also use the laboratory--frame, initial--meson momentum $k = {1 \over 2}
(\omega^2 - \mu^2)^{1/2}$ and energy $\omega$, related to the c.m. total
energy via $\omega = (w^2 - m^2 - \mu^2)/2m$, and, besides $t$, the two other
Mandelstam variables $s = w^2$ and $u$, the square of the c.m. total energy
for the crossed channel $\bar a_2(0^-) + B_1({1\over2}^+) \to \bar a_1(0^-) +
B_2({1\over2}^+)$, obeying {\sl on the mass shell} the indentity $s + t + u =
m^2 + m'^2 + \mu^2 + \mu'^2$.

In terms of the amplitudes $G$ and $H$ the c.m. differential cross section
for an unpolarized target takes the simple form
$$ {{d \sigma} \over {d \Omega}} = {1 \over 2} \sum_{\alpha,\beta} \vert
T_{\alpha\beta} \vert^2 = \vert G \vert^2 + \vert H \vert^2 \ . \eqno(13) $$

The other observable accessible at DA$\Phi$NE, in the strangeness--exchange
processes $\bar K N \to \pi \Lambda$, $\pi \Sigma$, is the polarization
$P_Y$ ($Y = \Lambda$, $\Sigma$) of the final hyperon, measurable through the
asymmetry $\alpha$ of their weak, nonleptonic decays $\Lambda \to \pi^- p$,
$\pi^0 n$ (both have an asymmetry $\alpha \simeq 0.64$), and $\Sigma^+ \to
\pi^0 p$ (which has $\alpha \simeq -0.98$), while there is little chance to
use the neutron decay modes $\Sigma^\pm \to \pi^\pm n$, which have $\alpha
\simeq \pm 0.068$; we have for these quantities
$$ P_Y \cdot ({{d \sigma} \over {d \Omega}}) = 2\ \hbox{\rm Im} \ (G \cdot
H^*) \ . \eqno(14) $$

For an $(S+P)$--wave parametrization (adequate at such low momenta), while the
integrated cross sections depend only quadratically on the $P$--waves, both
the first Legendre coefficients of the differential cross sections
$$ L_1 = {1 \over 2} \int_{-1}^{+1} \cos \theta \ ({{d \sigma} \over {d
\Omega}}) \ d \cos \theta = {2 \over 3} \ \hbox{\rm Re}\ [T_{0+} \cdot (2
T_{1+} + T_{1-})^* + \dots ] \eqno(15)$$
and the polarizations
$$ P_Y \cdot ({{d \sigma} \over {d \Omega}}) = 2 \ \hbox{\rm Im}\ [T_{0+}
\cdot (T_{1+} - T_{1-})^* + 3 T_{1-} \cdot T_{1+}^* \cos \theta + \dots ] \sin
\theta \eqno(16) $$
are essentially linear in the $P$--wave contributions, and give complementary
information on these latter. It is perhaps not useless to remind that the
low--statistics experiments performed only up to the late seventies have been
able to put only rather generous upper bounds$^{11}$ on these parameters for
the hyperon production channels.

We shall devote the last part of this section to show why this absence of
{\sl direct} information on the low--energy $P$--waves has been a serious
shortcoming for $\bar K N$ amplitude analyses. Remember that from {\sl
production} experiments we know that the $I = 1$, $S = -1$ $T_{1+}$ partial
wave resonates {\sl below} threshold at a c.m. energy around $w = 1385\ MeV$,
the mass of the isovector member of the $J^P = {3\over2}^+$ decuplet.

For any analytical extrapolation purpose, one has to turn from the Pauli
amplitudes $G$ and $H$ to the invariant amplitudes $A(s,t)$ and $B(s,t)$,
defined in term of four--component Dirac spinors as
$$ 2 \pi w \ T_{\alpha\beta} = \bar u_\alpha(p') [A(s,t) + B(s,t) \cdot
\gamma^\mu Q_\mu ] u_\beta(p) \ , \eqno(17) $$
where $Q = {1\over2} (q + q')$, the average between incoming-- and
outgoing--meson c.m. four--momenta: these amplitudes obey simple crossing
relations and are free of kinematical singularities, so that they are the
ones to be used, rather than $G$ and $H$. It is also customary to use the
combination $D(\nu,t) = A(\nu,t) + \nu \cdot B(\nu,t)$, where $\nu = (s - u)
/ 2(mm')^{1/2}$, which has the same properties as $A(\nu,t)$ under crossing
and for elastic scattering obeys the optical theorem in the simple form
$$ \hbox{\rm Im} \ D(\nu,t=0) = k \cdot \sigma_{tot} \ , \eqno(18) $$
where of course all electromagnetic effects must be subtracted on both sides.

$A$ and $B$ can be rewritten in terms of $G$ and $H$ and reexpressed
through the partial waves $T_{\ell\pm}$ by projecting eq. (17)
on the different spin states: for elastic scattering one gets
$$ A(\nu,t) = {{4 \pi} \over {E + m}} \{ (w + m) G(w,\theta) +$$
$$+ [(E + m)^2 (w -
m) + ({1\over2} t + E^2 - m^2) (w + m) ] \cdot{{H(w,\theta)} \over {q^2 \sin
\theta}} \} \ , \eqno(19)$$
and
$$ B(\nu,t) = {{4 \pi} \over {E + m}} \{ G(w,\theta) - [(E + m)^2 - {1\over2}
t - E^2 + m^2 ] {{H(w,\theta)} \over {q^2 \sin \theta}} \} \ . \eqno(20)$$

The amplitudes become, leaving out $D$-- and higher waves,
$$ D(\nu,0) = {{4 \pi w} \over m} [T_{0+} + 2 T_{1+} + T_{1-} + \dots ]
\eqno(21) $$
and
$$ B(\nu,0) = {{4 \pi w} \over {m q^2}} [(E - m) T_{0+} - 2 (2m - E) T_{1+} +
(E + m) T_{1-} + \dots ] \ . \eqno(122) $$
Introducing the (complex) scattering lengths $a_{\ell\pm}$ and (complex)
effective ranges $r_{\ell\pm}$ one can expand up to $O(q^2)$, and obtain
for the forward $D$ amplitudes close to threshold,
$$ D(q,0) = 4 \pi (1 + {\mu \over m}) \{ a_{0+} + i a_{0+}^2 q + [2 a_{1+} +
a_{1-} - (a_{0+} + {1\over2} r_{0+}) a_{0+}^2 - {a_{0+} \over {2 m \mu}}] q^2
+ $$
$$ + \dots \} \ , \eqno(23) $$
dominated by the $S$--waves, while for the $B$ amplitudes the same
approximations give
$$ B(q,0) = {{2 \pi} \over m} (1 + {\mu \over m}) [a_{0+} - 4 m^2 (a_{1+} -
a_{1-}) + i a_{0+}^2 q + \dots] \ , \eqno(24) $$
where the factor $4 m^2 \simeq 90\ fm^{-2}$ enhances the
low--energy P--waves (virtually unkown), rendering
practically useless the unsubtracted dispersion relation for the
better converging $B$ amplitudes, so important for the $\pi N$ case
in fixing accurately the values
of the coupling constant $f^2$ and of the S--wave scattering lengths$^{14}$.

\vglue 0.6truecm
\noindent{\bf 4. Impact of DA$\Phi$NE on baryon spectroscopy: the states
$\Lambda$(1405) and $\Sigma$(1385). }

At low momenta, comparable to those of the kaons from DA$\Phi$NE, we
have data from low--statistics experiments, mostly hydrogen
bubble--chamber ones on $K^- p$ (and $K^-d$) interactions$^{11,16}$
(dating from the early sixties trough the late seventies),
plus scant data from $K_L$
interactions and $K_S$ regeneration on hydrogen$^{17}$.

The inelastic channels, open at a laboratory energy
$\omega = {1\over2} M_\phi$
(for $K^\pm$'s the value of $\omega$ at the interaction point has to include
ionization energy losses as well), are the two--body ones $\pi\Lambda$ and
$\pi\Sigma$ (in all possible charge states), plus the three--body one
$\pi\pi\Lambda$ for $K^-$ or $K_L$ interacting with nucleons:
$K^+$--initiated processes are (apart from charge exchange) purely
elastic in this energy region.

For interactions in hydrogen, the c.m. energy is limited by
momentum conservation to the initial one,
equal (neglecting energy losses) to $w = (m_p^2 + \mu_K^2 + m_p
M_\phi)^{1/2}$, or $1442.4\ MeV$ for incident $K^\pm$'s and $1443.8\ MeV$
for incident $K_L$'s. Energy losses for charged kaons can be exploited
(using the inner parts of the detector as a moderator) to explore $K^- p$
interactions in a {\sl limited} momentum range, down to the charge--exchange
threshold at $w = 1437.2\ MeV$, corresponding to a $K^-$ laboratory momentum
of about $90\ MeV/c$.

For interactions in nuclei, momentum can be carried away by spectator
nucleons, and the inelastic channels can be explored down to threshold.
The possibility of reaching energies below the $\bar K N$ threshold allows
exploration of the unphysical region, containing two resonances, the $I = 0$,
$S$--wave $\Lambda(1405)$ and the $I = 1$, $J^P = {3\over2}^+$ $P$--wave
$\Sigma(1385)$, observed mostly in production experiments (and, in the first
case, in very limited statistics ones$^{13}$): the information on their
couplings to the $\bar K N$ channel relies entirely on extrapolations of the
low--energy $\bar K N$ data. The coupling of the $\Sigma(1385)$ to the
$\bar K N$ channel, for instance, can be determined via forward dispersion
relations involving the total sum of data collected at $t \simeq 0$, but still
with uncertainties which are, {\sl at their best}, still of the order of 50\%
of the flavour--$SU(3)$ symmetry prediction$^{18}$; as for the
$\Lambda(1405)$, even its spectroscopic classification is an open problem,
{\sl vis--\`a--vis} the paucity and (lack of) quality of the best available
data$^{19}$.

A formation experiment on {\sl bound} nucleons, in an (almost) $4 \pi$
apparatus with good efficiency and resolution for low--momentum $\gamma$'s
(such as KLOE$^{20}$), can measure a channel such as $K^- p \to \pi^0
\Sigma^0$ (above threshold), or $K^- d \to \pi^0 \Sigma^0 n_s$ (both above and
below threshold), which is pure $I = 0$: up to now all analyses on the
$\Lambda(1405)$ have been limited to charged channels$^{13}$, and assumed the
$I = 1$ contamination to be either negligible or smooth and non--interfering
with the resonance signal. Since the models proposed for the $\Lambda(1405)$
differ mostly in the details of the resonance shape, rather than in its
couplings, and it is precisely the shape which could be changed even by a
moderate interference with an $I = 1$ background, such measurements would be
decisive. Having in the same apparatus and at almost the same energy {\sl
tagged} $K^-$ and $K_L$ produced at the same point, one can further separate
$I=0$ and $I=1$ channels with a minimum of systematic uncertainties, by
measuring all channels $K_L p \to \pi^0 \Sigma^+$, $\pi^+ \Sigma^0$ and $K^- p
\to \pi^- \Sigma^+$, $\pi^+ \Sigma^-$, besides, of course, the
above--mentioned, pure $I=0$, $K^- p \to \pi^0 \Sigma^0$ one.

Another class of inelastic processes which are expected to be produced,
at a much smaller rate, by DA$\Phi$NE's kaons are the radiative capture
processes $K^- p \to \gamma \Lambda$, $\gamma \Sigma^0$ and $K_L p \to \gamma
\Sigma^+$ (both in hydrogen and deuterium), and $K^- n \to \gamma \Sigma^-$
and $K_L n \to \gamma \Lambda$, $\gamma \Sigma^0$ (only in deuterium).
Up to now
only searches for photons emitted after stops of $K^-$'s in liquid hydrogen
and deuterium have been performed with some success: the spectra are
dominated by photons from unreconstructed $\pi^0$ and $\Sigma^0$
decays$^{21}$, and separating the signals from this background poses serious
difficulties, since only the photon line from the $\gamma \Lambda$ final state
falls just above the endpoint of the photons from $\pi^0$ decays in the $\pi^0
\Lambda$ final state, while that from $\gamma \Sigma^0$ falls right on top of
the latter. Indeed these experiments were able to produce only an estimate of
the respective branching ratios$^{21}$.

The $4\pi$ geometry possible at DA$\Phi$NE, combined with the ``transparency''
of a KLOE--like apparatus$^{20}$, its high efficiency for photon detection and
its good resolution for spatial reconstruction of the events, should make
possible the full identification of the final states and therefore the
measurement of the absolute cross sections for these processes, although in
flight and not at rest.

Data$^{21}$ are presently indicating branching ratios around $0.9 \times
10^{-3}$ for $K^- p \to \gamma \Lambda$ and $1.4 \times 10^{-3}$ for $K^- p
\to \gamma \Sigma^0$, with errors of the order of 15\% on both: most
models$^{23}$ give the first rate larger than the second, with both values
consistently higher than the observed ones. Only a cloudy--bag--model$^{24}$
exhibits the trend appearing (although only at a $2\sigma$--level, and
therefore waiting for confirmation by better data) from the first experimental
determinations, but this is the only respect in which it agrees with the data,
still giving branching ratios larger than observations by a factor two.

Data are also interpretable in terms of $\Lambda(1405)$ electromagnetic
transition moments$^{22}$: this interpretation is clearly sensitive to the
interference between the decay of this state and all other contributions. An
extraction of the $\Lambda(1405)$ moments freer of these uncertainties would
require measurements of $\gamma \Lambda$ and $\gamma \Sigma$ (if possible, in
different charge states) over the unphysical region, using (gaseous) deuterium
or helium as a target. Rates are expected to be of the order of $10^4\
events/y$ only, but such a low rate would correspond to better statistics than
those of the best experiment performed on the $\Lambda(1405) \to \pi \Sigma$
decay spectrum$^{13}$.

\vglue 0.6truecm
\noindent{\bf 5. Description of coupled $\bar KN$, $\pi Y$ channels: the
K--matrix. }

A description of the low--energy $\bar K N$ partial waves must couple the
two--body inelastic channels to each other and to the elastic one: the
three--body channel $\pi \pi \Lambda$ is expected to be suppressed, for $J^P
= {1\over2}^-$, by the angular momentum barrier, but it could contribute
appreciably to the $I=0$, $J^P={1\over2}^+$ $P$--wave, due to the strong
final--state interaction of two pions in an $I=0$ $S$--wave. Most
bubble--chamber experiments were unable to fully reconstruct events at the
lowest momenta, and therefore assumed all {\sl directly} produced $\Lambda$'s
to come from the $\pi \Lambda$ channel alone, neglecting the small $\pi \pi
\Lambda$ contribution altogether.

The appropriate formalism is to introduce a K--matrix description (sometimes
it is convenient to use, instead of the K--matrix, its inverse, {\sl a.k.a.}
the M--matrix), defined in the isospin eigenchannel notation as
$$ \hbox{\bf K}_{\ell\pm}^{-1} = \hbox{\bf M}_{\ell\pm} =
\hbox{\bf T}_{\ell\pm}^{-1} + i\ \hbox{\bf Q}^{2\ell +1}\ , \eqno(25)$$
for both $I=0, 1$ $S$--waves (and perhaps also for the four $P$--waves
as well). The K--matrices, assuming $SU(2)$ symmetry, describe the $S$--wave
data at a given energy in terms of {\sl nine} real parameters (six for $I=1$
and three for $I=0$), while the experimentally accessible processes are
described, with pure $S$--waves and in the same symmetry limit, by only {\sl
six independent} parameters, for which one can choose the two (complex)
amplitudes $A_{0,1}$ for the elastic channel, the phase difference $\phi$
between the $I=0$ and $I=1$ $\pi \Sigma$ production amplitudes, and the ratio
$\epsilon$ between the $\pi \Lambda$ production cross section and that for
total hyperon production in an $I=1$ state$^{25}$.

Thus a single--energy measurement does not allow a complete determination of
the K--matrix elements. For a precise determination of the $S$--waves one
should also subtract out the $P$--wave contributions to the integrated cross
sections
$$ \sigma = 4\pi L_0 = 2\pi \int_{-1}^{+1} ({{d\sigma}\over{d\Omega}})\
d\cos\theta = 4\pi [ \vert T_{0+} \vert^2 + 2 \vert T_{1+}\vert^2 + \vert
T_{1-}\vert^2 + \dots ] \ , \eqno(26)$$
which could be obtained either from $L_1$ alone for the elastic and
charge--exchange channels, or from both $L_1$ and $P_Y$ for the hyperon
production channels. None of these quantities has been measured up to now:
the TST Collaboration tried to extract $L_1$ from some of their
low--statistics data, but found results consistent with zero within their
obviously very large errors$^{11}$. At the same level of accuracy, one should
also be able to isolate out the $\pi \pi \Lambda$ channel as well. Note that
an accurate analysis has also to include complete e.m.
corrections$^{15,26}$: up to now all $\bar KN$ analyses have relied on the
old, approximate formul\ae\ derived by Dalitz and Tuan for a pure $S$--wave
scattering$^{27}$.

To fix the redundant K--matrix parameters different ways have been tried: some
authors have used the data on the shape of the $\pi \Sigma$ spectrum from
production experiments$^{28}$, others have constrained the amplitudes in the
unphysical region by imposing consistency with forward dispersion relations
for both $K^\pm p$ and $K^\pm n$ elastic--scattering $D$ amplitudes$^{29}$,
relying on the accurate total--cross--section data at higher energies. More
recently, some attempts have been made to combine both constraints into a
global analysis, but with no better results than each of them taken
separately$^{30}$.

Unfortunately, neither of these methods has been very powerful because of the
low statistics of the $\pi \Sigma$--production data on one side, and on the
other because of the need to use for the dispersion relations the often
inaccurate information (and particularly so for the $K^\pm n$ amplitudes) on
the real--to--imaginary--part ratios.

We list below the constant K--matrices found by Chao {\sl et al.} using the
first method$^{28}$, which did not include the TST Collaboration data, and the
more complex parametrization found by A.D. Martin using the second$^{29}$, and
including the preliminary TST data. Note that to describe the data for $I=0$
both above and below threshold A.D. Martin had to introduce a
constant--effective--range M--matrix ${\bf M}^{(0)} = ({\bf K}^{(0)})^{-1} =
{\bf A} + {\bf R} k^2$, so that
to make the two analyses comparable we list separately his threshold K--matrix
values.

\vglue 0.3truecm
\centerline{\bf Table I}
\vglue 0.3truecm
\hrule
$$\vbox{\halign{\hfil#\ \hfil&\hfil#\ \hfil&\hfil#\ \hfil&\hfil#\ \hfil\cr
\sl Chao et al.&&\sl A. D. Martin&\cr
&&&\cr
$K_{NN}^{(0)} = -1.56fm$&$A_{NN} = -0.07fm^{-1}$&$R_{NN} = +0.18fm$&
$K_{NN}^{(0)}(0) = -1.65fm$\cr
$K_{N\Sigma}^{(0)} = -0.92fm$&$A_{N\Sigma} = -1.02fm^{-1}$&$R_{N\Sigma} =
+0.19fm$&$K_{N\Sigma}^{(0)}(0) = +0.16fm$\cr
$K_{\Sigma\Sigma}^{(0)} = +0.07fm$&$A_{\Sigma\Sigma} = +1.94fm^{-1}$&
$R_{\Sigma\Sigma} = -1.09fm$&$K_{\Sigma\Sigma}^{(0)}(0) = -0.15fm$\cr
&&&\cr
$K_{NN}^{(1)} = +0.76fm$&&&$K_{NN}^{(1)} = +1.07fm$\cr
$K_{N\Sigma}^{(1)} = -0.97fm$&&&$K_{N\Sigma}^{(1)} = -1.32fm$\cr
$K_{N\Lambda}^{(1)} = -0.66fm$&&&$K_{N\Lambda}^{(1)} = -0.30fm$\cr
$K_{\Sigma\Sigma}^{(1)} = +0.86fm$&&&$K_{\Sigma\Sigma}^{(1)} = +0.27fm$\cr
$K_{\Sigma\Lambda}^{(1)} = +0.51fm$&&&$K_{\Sigma\Lambda}^{(1)} = +1.54fm$\cr
$K_{\Lambda\Lambda}^{(1)} = +0.04fm$&&&$K_{\Lambda\Lambda}^{(1)} = -1.02fm$
\cr}}$$
\hrule
\vglue 0.8truecm

The table shows that there is considerable uncertainty even on the
$K^{(I)}_{NN}$ elements (the real parts of the corresponding scattering
lengths): the data have been re--analyzed by Dalitz {\sl et al.}$^{30}$, using
both sets of constraints with different weigths and different
parametrizations, and yielding a variety of fits, all of them of about the
same quality of, but none of them improving very much over, the above ones.

To further highlight the difficulties met in fitting the data, we point out
that A.D. Martin himself$^{29}$ found that including in his analysis a
$\Sigma(1385)$ resonance with the width given by production experiments and
the coupling to the $\bar K N$ channel dictated by flavour--$SU(3)$ symmetry,
was worsening rather than improving the results obtained neglecting it
altogether. He proposes therefore to consider the $\Sigma$ Born--term
contribution a superposition of the former and of that of the $P$--wave
resonance: a rather unsavoury situation, considering the different $J^P$ of
the two states, which may raise questions about the applicability of his
analysis away from $t \simeq 0$. An analogous superposition is considered in
$K^\pm p$ dispersion relations, where one can not separate the $\Sigma$-- from
the $\Lambda$--pole, but here the two contribute to the same partial wave,
and the $\Sigma$--pole can be extracted independently from data on $K^\pm n$
scattering and $K_S$ regeration on protons$^{31,32}$.

In the analysis of the low--energy data collected in the past, a further
difficulty comes from the large momentum spread of the low--energy
kaon beams, for $K^\pm$'s because of the degrading in a moderator of the
higher--energy beams needed to transport the kaons away from their production
targets, for $K_L$'s because of the large apertures needed to achieve
satisfactory rates in the targets (typically bubble chambers): this made
unrealistic the proposals (advanced from the early seventies) of better
determining the low--energy K--matrices by studying the behaviour of the
cross sections for $K^-p$--initiated processes at the $\bar K^0n$
charge--exchange threshold$^{10}$. The high momentum resolution
available at DA$\Phi$NE will instead make such a goal a realistically
achievable one.

In this case $SU(2)$ can no longer be assumed to be a symmetry of
the amplitudes: under the (reasonable) assumption that the forces are
still $SU(2)$--symmetric, one can however retain the previous K--matrix
formalism, but no longer decouple the different isospin eigenchannels$^{33}$.
Introducing the orthogonal matrix {\bf R}, which transforms the six isospin
eigenchannels for $\bar KN$ ($I=0, 1$), $\pi\Lambda$ ($I=1$ only) and $\pi\
\Sigma$ ($I=0, 1, 2$) into the six charge--states $K^-p$, $\bar K^0n$, $\pi^0
\Lambda$, $\pi^-\Sigma^+$, $\pi^0\Sigma^0$ and $\pi^+\Sigma^-$, and the
diagonal matrix {\bf Q}$_c$ of the c.m. momenta for these latter, one can
rewrite the T--matrix in the isospin--eigenchannel space for the $S$--waves as
$$ \hbox{\bf T}_I^{-1} = \hbox{\bf K}_I^{-1} - i \hbox{\bf R}^{-1} \hbox{\bf
Q}_c \hbox{\bf R} \ , \eqno(27)$$
where {\bf K}$_I$ is a box matrix with zero elements between channels of
different isospin, and {\bf R}$^{-1}$ {\bf Q}$_c$ {\bf R} is of course no
longer diagonal.

Apparently this involves one more parameter, since it
also contains the element $K_{\Sigma\Sigma}^{(2)}$: in practice, if one is
interested in the cross sections only in the neighbourhood of
the $\bar KN$ charge--exchange threshold, one can take the
c.m. momenta in the three $\pi\Sigma$ channels as equal and decouple the
$I=2$, $\pi\Sigma$ channel from the $I=0,\ 1$ ones, since the ``rotated''
matrix {\bf R}$^{-1}${\bf Q}$_c${\bf R}
has now only two non--zero, off-diagonal elements, equal to
${1\over2} (q_0 - q_-)$ (the subscripts refer to the kaon charges),
between the $I=0$ and $I=1$ $\bar KN$ channels, the diagonal ones being the
same as in the $SU(2)$--symmetric case, if one substitutes for the
$\bar KN$ channel momentum $q$ the average over the two charge states
${1\over2} (q_0 + q_-)$. Indeed $K^{(2)}_{\Sigma\Sigma}$ could only be
important for an accurate description of the $\pi\Sigma$ and $\pi\Lambda$ mass
spectra close to the $\pi\Sigma$ threshold.

\vglue 0.6truecm
\noindent{\bf 6. Outline of a theoretical program for a modern $KN$ amplitude
analysis. }

The measurements proposed for DA$\Phi$NE will provide data of the same
statistical quality now available {\sl only} for the $\pi N$ system:
theoretical tools for their analysis must thus be improved as well, to meet
the standards required by this, long awaited for, ``forward leap'' in $KN$
data. Since long, tools of just this level have been provided, for $\pi N$
amplitude analysis, by the so--called ``Karlsruhe--Helsinki collaboration''
headed over the years by Prof. G. H\"ohler$^{14,34}$: alas, their software
can not be straightforwardly imported to do $K N$ analyses, mainly because of
the complicated analytic structure of the low-energy $\bar KN$ amplitudes.

Much for the same reason, the dispersive treatment of Coulomb corrections,
developed at NORDITA$^{15,26}$ by Hamilton and collaborators, can not be
immediately transferred to the strange sector. It has to be recalled that old
data were always analyzed using, for these corrections, the approximate
formula of Dalitz and Tuan$^{27}$, which considers only a pure--$S$--wave
strong interaction, and furthermore might be inapplicable to a {\sl strongly
absorptive} interaction close to threshold$^{35}$.

Since the basic principles, on which both approaches are based, have to
hold for the $\bar K N$ system as for the $\pi N$ one, it remains only to
work out the details of a partial--wave--analysis procedure, applicable to a
system strongly absorptive at threshold, and possessing {\sl ab--initio} the
following requirements:
i) consistency with both fixed--$t$ and partial--wave dispersion relations;
ii) crossing symmetry (and isotopic--spin symmetry as well, to describe {\sl
simultaneously} charge--exchange and regeneration data);
iii) analyticity in $t$ beyond the Lehman ellipses, with the {\sl correct}
low--mass, $t$--channel--cut discontinuities given by the $\pi\pi$ cut;
iv) a {\sl complete} treatment of radiative and Coulomb corrections.

The author and its collaborators (chiefly G. Violini of Universit\`a
della Calabria, Cosenza, and C.I.F., Bogot\'a, and R.C. Barrett of
the University of
Surrey, Guildford) have in the past carried out parts of this program
(as many others have done, more or less at the same epoch), but only for
{\sl limited purposes}, such as extrapolations either to the hyperon
poles$^{31,36}$ or to the Cheng--Dashen point$^{37}$, or studies of Coulomb
effects$^{35}$ and radiative capture at threshold$^{24}$:
what remains to be done is a merging together of all these techniques into a
``global'' analysis, on which work is presently under way.

In this perspective we have advanced the proposal to I.N.F.N. for a
program of extensive collaboration, code--named KILN (for ``Kaon Interactions
at Low energies with Nucleons''), which has already received an initial,
and thus limited, financial support. Participation in this
collaboration is highly welcome, and we take this
occasion for calling upon all theorists wich have been or wish to be active
in this still open and very much alive (despite greatly exaggerated rumors on
the contrary$^{38}$) field of particle physics.

\vglue 0.6truecm
\centerline{\bf ACKNOWLEDGEMENTS }

Exchanges of information with theorists and
experimentalists involved in the planning of experiments at both
machines, KAON and DA$\Phi$NE, is gratefully acknowledged by the author and
his collaborators.

\vfill\eject
\vglue 0.6truecm
\centerline{\bf REFERENCES }

\item{1.} See the Proceedings of the {\sl ``Workshop on Physics and
Detectors for DA$\Phi$NE''}, ed. by G. Pancheri (I.N.F.N., Frascati 1991).

\item{2.} M. Aguilar--Benitez, {\sl et al.} (Particle Data Group): {\sl
Phys. Lett.} {\bf B 239} (1990) 1.

\item{3.} P.M. Gensini and G. Violini: {\sl ``Workshop on Science at the
KAON Factory''}, ed. by D.R. Gill (TRIUMF, Vancouver 1991), Vol. II, p. 193;
P.M. Gensini: {\sl ``Workshop on Physics and Detectors for DA$\Phi$NE''}, ed.
by G. Pancheri (I.N.F.N., Frascati 1991), p. 453.

\item{4.} C.J. Batty and A. Gal: {\sl Nuovo Cimento} {\bf 102 A} (1989)
255; C.J. Batty: {\sl ``First Workshop on Intense Hadron Facilities and
Antiproton Physics''}, ed. by T. Bressani, F. Iazzi and G. Pauli (S.I.F.,
Bologna 1990), p. 117.

\item{5.} J. Beveridge: {\sl ``Workshop on Science at the KAON Factory''},
ed. by D.R. Gill (TRIUMF, Vancouver 1991), Vol. I, p. 19.

\item{6.} G. Vignola: {\sl ``Workshop on Physics and Detectors for
DA$\Phi$NE''}, ed. by G. Pancheri (I.N.F.N., Frascati 1991), p. 11.

\item{7.} T. Bressani: {\sl ``Workshop on Physics and Detectors for
DA$\Phi$NE''}, ed. by G. Pancheri (I.N.F.N., Frascati 1991), p. 475; {\sl
"Common Problems and Ideas of Modern Physics"}, ed. by T. Bressani, B. Minetti
and A. Zenoni (World Scientific, Singapore 1992), p. 222, and the talk
presented at this Folgaria Winter School.

\item{8.} See the Proceedings of the {\sl ``Workshop on Science at the KAON
Factory''}, ed. by D.R. Gill (TRIUMF, Vancouver 1991).

\item{9.} {\sl ``Proposal for a European Hadron Facility''}, ed. by J.F.
Crawford, report {\sl EHF--87--18} (Trieste--Mainz, may 1987).

\item{10.} See the discussion on this point by D.J. Miller: {\sl ``Proc. of
the Int. Conf. on Hypernuclear and Kaon Physics''}, ed. by B. Povh (M.P.I.,
Heidelberg 1982), p. 215.

\item{11.} TST Collaboration: R.J. Novak, {\sl et al.: Nucl Phys.} {\bf B
139} (1978) 61; N.H. Bedford, {\sl et al.: Nukleonika} {\bf 25} (1980) 509; M.
Goossens, G. Wilquet, J.L. Armstrong and J.H. Bartley: {\sl ``Low and
Intermediate Energy Kaon--Nucleon Physics''}, ed. by E. Ferrari and G. Violini
(D. Reidel, Dordrecht 1981), p. 131; J. Ciborowski, {\sl et al.: J. Phys.}
{\bf G 8} (1982) 13; D. Evans, {\sl et al.: J. Phys.} {\bf G 9} (1983) 885; J.
Conboy, {\sl et al.: J. Phys} {\bf G 12} (1986) 1143. A good description of
the experiment is in D.J. Miller, R.J. Novak and T. Tyminiecka: {\sl ``Low and
Intermediate Energy Kaon-Nucleon Physics''}, ed by E. Ferrari and G. Violini
(D. Reidel, Dordrecht 1981), p. 251.

\item{12.} E. Piasetzky: {\sl Nuovo Cimento} {\bf 102 A} (1989) 281.

\item{13.} R.J. Hemingway: {\sl Nucl. Phys.} {\bf B 253} (1985) 742. Older
data are even poorer in statistics: see ref. 27 for a comparison. See also,
for formation on bound nucleons, B. Riley, I.T. Wang, J.G. Fetkovich and J.M.
McKenzie: {\sl Phys. Rev.} {\bf D 11} (1975) 3065.

\item{14.} For conventions and kinematical notations we have adopted the
same as: G. H\"ohler, F. Kaiser, R. Koch and E. Pietarinen: {\sl ``Handbook of
Pion--Nucleon Scattering''} (Fachinformationszentrum, Karlsruhe 1979), and
{\sl ``Landolt--B\"orn-stein, New Series, Group I, Vol. 9b''}, ed. by H.
Schopper (Springer--Verlag, Berlin 1983), which have become a ``standard''
for describing $\pi N$ scattering.

\item{15.} B. Tromborg, S. Waldenstr\"om and I. \O verb\o : {\sl Ann. Phys.
(N.Y.)} {\bf 100} (1976) 1; {\sl Phys. Rev.} {\bf D 15} (1977) 725; {\sl Helv.
Phys. Acta} {\bf 51} (1978) 584.

\item{16.} $K^\pm p$ data: W.E. Humphrey and R.R. Ross: {\sl Phys. Rev.}
{\bf 127} (1962) 1; G.S. Abrams and B. Sechi--Zorn: {\sl Phys. Rev.} {\bf 139}
(1965) B454; M. Sakitt, {\sl et al.: Phys Rev} {\bf 139} (1965) B719; J.K.
Kim: Columbia Univ. report {\sl NEVIS--149} (New York 1966) and {\sl Phys.
Rev. Lett.}
{\bf 14} (1970) 615; W. Kittel, G Otter and I. Wa\v{c}ek: {\sl Phys. Lett}
{\bf 21} (1966) 349; D. Tovee, {\sl et al.: Nucl. Phys.} {\bf B 33} (1971)
493; T.S. Mast, {\sl et al.: Phys. Rev.} {\bf D 11} (1975) 3078 and {\bf D 14}
(1976) 13; R.O. Bargenter, {\sl et al.: Phys. Rev.} {\bf D 23} (1981) 1484.
$K^-d$ data: R. Armenteros, {\sl et al.: Nucl. Phys.} {\bf B 18} (1970) 425.

\item{17.} $K_Lp$ data: J.A. Kadyk, {\sl et al.: Phys. Lett.} {\bf 17}
(1966) 599 and report {\sl UCRL--18325} (Berkeley 1968); R.A. Donald, {\sl et
al.:
Phys. Lett.} {\bf 22} (1966) 711; G.A. Sayer, {\sl et al.: Phys. Rev.} {\bf
169} (1968) 1045.

\item{18.} G.C. Oades: {\sl Nuovo Cimento} {\bf 102 A} (1989) 237.

\item{19.} A.D. Martin, B.R. Martin and G.G. Ross: {\sl Phys. Lett.} {\bf B
35} (1971) 62; P.N. Dobson jr. and R. McElhaney: {\sl Phys. Rev.} {\bf D 6}
(1972) 3256; G.C. Oades and G. Rasche: {\sl Nuovo Cimento} {\bf 42 A} (1977)
462; R.H. Dalitz and J.G. McGinley: {\sl ``Low and Intermediate Energy
Kaon--Nucleon Physics''}, ed. by E. Ferrari and G. Violini (D. Reidel,
Dordrecht 1981), p. 381, and the Ph.D. thesis by McGinley (Oxford Univ. 1979);
G.C. Oades and G. Rasche: {\sl Phys. Scr.} {\bf 26} (1982) 15; J.P. Liu: {\sl
Z. Phys.} {\bf C 22} (1984) 171; B.K. Jennings: {\sl Phys. Lett.} {\bf B 176}
(1986) 229; P.B. Siegel and W. Weise: {\sl Phys. Rev.} {\bf C 38} (1988) 2221;
R.H. Dalitz and A. Deloff: {\sl J. Phys.} {\bf G 17} (1991) 289.

\item{20.} A. Aloisio, {\sl et al.} (The KLOE Collaboration): {\sl ``KLOE
-- A General Purpose Detector for DA${\mit \Phi}$NE''}, report {\sl
LNF--92/019 (IR)} (Frascati, April 1992).

\item{21.} B.L. Roberts: {\sl Nucl Phys} {\bf A 479} (1988) 75c; B.L.
Roberts, {\sl et al.: Nuovo Cimento} {\bf 102 A} (1989) 145; D.A. Whitehouse,
{\sl et al.: Phys. Rev. Lett.} {\bf 63} (1989) 1352.

\item{22.} See the review by J. Lowe: {\sl Nuovo Cimento} {\bf 102 A}
(1989) 167.

\item{23.} J.W. Darewich, R. Koniuk and N. Isgur: {\sl Phys. Rev.} {\bf D
32} (1985) 1765; H. Burkhardt, J. Lowe and A.S. Rosenthal: {\sl Nucl Phys.}
{\bf A 440} (1985) 653; R.L. Workman and H.W. Fearing: {\sl Phys. Rev.} {\bf D
37} (1988) 3117; R.A. Williams, C.R. Ji and S. Cotanch: {\sl Phys. Rev.} {\bf
D 41} (1990) 1449; {\sl Phys. Rev.} {\bf C 43} (1991) 452; H. Burkhardt and J.
Lowe: {\sl Phys. Rev.} {\bf C 44} (1991) 607. For radiative capture on
deuterons (and other light nuclei), see: R.L. Workman and H.W. Fearing: {\sl
Phys. Rev.} {\bf C 41} (1990) 1688; C. Bennhold: {\sl Phys. Rev.} {\bf C 42}
(1990) 775.

\item{24.} Y.S. Zhong, A.W. Thomas, B.K. Jennings and R.C. Barrett: {\sl
Phys. Lett.} {\bf B 171} (1986) 471; {\sl Phys. Rev.} {\bf D 38} (1988) 837
(which corrects a numerical error contained in the previous paper).

\item{25.} See the review presented by B.R. Martin at the 1972 Ba\v{s}ko
Polje Int. School, published in: {\sl ``Textbook on Elementary Particle
Physics. Vol. 5: Strong Interactions''}, ed by M. Nikoli\v{c} (Gordon and
Breach, Paris 1975).

\item{26.} J. Hamiltom, I. \O verb\o\ and B. Tromborg: {\sl Nucl. Phys.}
{\bf B 60} (1973) 443; B. Tromborg and J. Hamilton: {\sl Nucl. Phys.} {\bf B
76} (1974) 483; J. Hamilton: {\sl Fortschr. Phys.} {\bf 23} (1975) 211.

\item{27.} R.H. Dalitz and S.F. Tuan: {\sl Ann. Phys. (N.Y.)} {\bf 10}
(1960) 307.

\item{28.} Y.A. Chao, R. Kr\"amer, D.W. Thomas and B.R. Martin: {\sl Nucl.
Phys.} {\bf B 56} (1973) 46.

\item{29.} A.D. Martin: {\sl Phys. Lett.} {\bf B 65} (1976) 346; {\sl ``Low
and Intermediate Energy Kaon-Nucleon Physics''}, ed. by E. Ferrari and G.
Violini (D. Reidel, Dordrecht 1981), p. 97; {\sl Nucl. Phys.} {\bf B 179}
(1981) 33.

\item{30.} R.H. Dalitz, J. McGinley, C. Belyea and S. Anthony: {\sl ``Proc.
of the Int. Conf. on Hypernuclear and Kaon Physics''}, ed. by B. Povh (M.P.I.,
Heidelberg 1982), p. 201.

\item{31.} G.K. Atkin, B. Di Claudio, G. Violini and N.M. Queen: {\sl Phys.
Lett.} {\bf B 95} (1980) 447; {\sl ``Low and Intermediate Energy Kaon--Nucleon
Physics''}, ed. by E. Ferrari and G. Violini (D. Reidel, Dordrecht 1981), p.
131.

\item{32.} J. Antol\'\i{n}: {\sl Phys. Rev.} {\bf D 43} (1991) 1532.

\item{33.} G.C. Oades and G. Rasche: {\sl Phys. Rev.} {\bf D 4} (1971)
2153; H. Zimmermann: {\sl Helv. Phys. Acta} {\bf 45} (1973) 1117; G. Rasche
and W.S. Woolcock: {\sl Fortschr. Phys.} {\bf 25} (1977) 501.

\item{34.} R. Koch: {\sl ``Low and Intermediate Energy Kaon-Nucleon
Physics''}, ed. by E. Ferrari and G. Violini (D. Reidel, Dordrecht 1981),
p. 1; see also the update by G. H\"ohler: {\sl $\pi N$ Newsletter} {\bf 2}
(1990) 1.

\item{35.} P.M. Gensini and G.R. Semeraro: {\sl ``Perspectives on
Theoretical Nuclear Phy-sics''}, ed. by L. Bracci, {\sl et al.} (ETS Ed., Pisa
1986), p. 91.

\item{36.} B. Di Claudio, G. Violini and N.M. Queen: {\sl Nucl. Phys.} {\bf
B 161} (1979) 238; G.K. Atkin, B. Di Claudio, G. Violini, J.E. Bowcock and
N.M. Queen: {\sl Z. Phys.} {\bf C 7} (1981) 249.

\item{37.} P.M. Gensini: {\sl J. Phys. G} {\bf 7} (1981) 1177; {\sl Nuovo
Cimento} {\bf 84 A} (1984) 203.

\item{38.} The same also happened to Mark Twain, from which we took the
liberty of borrowing the pun.

\bye